\documentclass[prd,preprint,superscriptaddress,amsmath,amssymb,nofootinbib]{revtex4}
\usepackage{graphicx}
\usepackage{dcolumn}
\usepackage{bm}
\usepackage{amssymb}
\usepackage{amsmath}
\usepackage[compat=1.0.0]{tikz-feynman}
\usepackage{epsfig}    
\usepackage{color}
\usepackage{slashed}
\usepackage{hhline}
\usepackage{bbm}
\usepackage{tikz}
\usepackage{tikz-feynman}
\usetikzlibrary{calc}
\tikzfeynmanset{compat=1.1.0}

\def\be{\begin{equation}}
\def\ee{\end{equation}}
\newcommand{\bea}{\begin{eqnarray}}
\newcommand{\eea}{\end{eqnarray}}
\newcommand{\nn}{\nonumber}

\begin{document}

\title{Radiative double inverse seesaw and dark matter in an alternative gauged $U(1)_{B-L}$ model}

\author{ Hiroshi Okada}
\email{hiroshi3okada@htu.edu.cn}
\affiliation{Department of Physics, Henan Normal University, Xinxiang 453007, China}

\author{Labh Singh}
\email{sainilabh5@gmail.com}
\affiliation{Independent Researcher, Solan 174101, India}

\date{\today}

\begin{abstract}
\noindent We propose a concrete theoretical framework for the double inverse seesaw mechanism within an alternative gauged $U(1)_{B-L}$ model, where the tiny neutrino masses emerge radiatively at the loop level in accordance with the 't Hooft naturalness criterion. By assigning non-universal $B-L$ charges of $(-4, -4, 5)$ to the right-handed neutrinos, we introduce vector-like fermions and two inert singlet scalars that circulate in the loop to generate the required mass terms. The spontaneous breaking of the $U(1)_{B-L}$ symmetry leaves a remnant $Z_2$ symmetry, which naturally stabilizes these new particles as dark matter (DM) candidates. We systematically investigate both fermionic and bosonic DM scenarios, assuming their interactions are predominantly mediated by the $B-L$ gauge boson ($Z'$). Our comprehensive analysis of the relic density, direct detection, and collider bounds reveals that the fermionic DM scenario is strongly favored. In contrast, the bosonic DM via the $Z'$ portal is severely constrained and largely ruled out by the latest direct detection experiments such as LZ, PandaX-4T, and XENONnT.

 \end{abstract}
\maketitle

\section{Introduction}
\noindent Elucidating the mechanism behind tiny neutrino masses is a central challenge in high-energy physics phenomenology. 
Although there are a vast amount of literature on how to induce such small masses,
the inverse seesaw mechanism~\cite{Mohapatra:1986bd, Wyler:1982dd} is one of the most attractive low-scale scenarios for neutrino mass generation.
This is because the smallness of neutrino masses is directly linked to a naturally small symmetry-breaking parameter that is known as the  ``'t Hooft's naturalness criterion"~\cite{tHooft:1980xss}~\footnote{The essential idea is that a tiny mass parameter should recover a symmetry if  the mass parameter goes to zero.}. 
An intriguing extension of this idea is the double inverse seesaw~\cite{CentellesChulia:2020dfh}, in which the light neutrino mass is suppressed by two insertions of the small symmetry-breaking parameter instead of one. This additional suppression enlarges the viable parameter space, allowing comparatively larger symmetry-breaking scales or lighter mediator masses without sacrificing naturally sized Yukawa couplings. Moreover, the double inverse seesaw constitutes the first member of a broader class of multiple inverse seesaw frameworks, offering a systematic approach to constructing neutrino mass models with increasingly enhanced suppressions while maintaining the naturalness of the underlying symmetry-breaking structure.
While the concept of technical naturalness in the sense of 't Hooft~\cite{tHooft:1980xss} provides a profound and conceptually natural foundation for understanding these mass hierarchies, it often lacks a concrete, predictive theoretical framework. To date, translating this naturalness argument into a robust and complete model has remained a significant challenge. 

\noindent In this paper, we aim to bridge this gap by providing a concrete theoretical framework for this concept, thereby establishing a more solid theoretical structure. Specifically, we postulate that the small mass scales are not generated at the tree level, but rather emerge radiatively at the loop level. This loop-induced mass generation naturally suppresses the mass scale, aligning perfectly with the 't Hooft naturalness criterion while offering a testable model building strategy.
To construct this model, we introduce an alternative gauged $U(1)_{B-L}$ symmetry~\cite{Ma:2014qra, Ma:2015mjd, Montero:2007cd} as the sole additional gauge symmetry. The particle content is kept minimal yet sufficient: in addition to the fermions required for the double inverse seesaw mechanism and the scalar boson responsible for the spontaneous breaking of the $U(1)_{B-L}$ symmetry, we introduce only vector-like fermions and two inert singlet scalars. These additional particles are specifically chosen to circulate in the loop and generate the desired mass terms radiatively~\cite{Nomura:2017vzp, Nomura:2017jxb, Geng:2017foe, Nomura:2017kih, Okada:2021nwo}.
A remarkable feature of this setup is its direct implication for dark matter (DM). Even after the spontaneous breaking of the $U(1)_{B-L}$ symmetry, a remnant discrete $Z_2$ symmetry remains unbroken. Consequently, the new particles running in the loop---the vector-like fermions and the inert singlet scalars---are automatically stabilized by this remnant symmetry and can serve as viable DM candidates. Within this framework, we systematically investigate both fermionic and bosonic DM scenarios.
To simplify the phenomenological analysis and focus on the core dynamics, we assume that the DM interactions are predominantly mediated by the $B-L$ gauge boson ($Z'$). Under this assumption, we evaluate the viability of the model by calculating the DM relic abundance and confronting it with current direct detection constraints and collider bounds. Our comprehensive analysis reveals that the fermionic DM scenario is generally favored over the bosonic one in simultaneously satisfying all cosmological and experimental constraints.

\noindent The remainder of this paper is organized as follows. In Sec.~\ref{sec:model}, we describe the model setup and the radiative mass generation mechanism. Sec.~\ref{sec:dm} is devoted to the discussion of the dark matter candidates, and presents our phenomenological analysis, including the relic density and experimental bounds.
Finally, we give our conclusions in Sec.~\ref{sec:con}.


\section{ Model setup and Phenomenologies}
\label{sec:model}
\noindent In this section, we introduce our model.
In the fermionic sector, we introduce two types of vector-like neutral particles $N$ and $\psi$, and the right-handed neutral particles $\nu_R$, where all these fields have three generations.~\footnote{The minimal number of the $N$ and $\psi$ can be two families, but we fix three families for completeness.}
$N$, $\psi$, and $\nu_R\equiv(\nu_{R_\rho},\nu_{R_3})^T$ are respectively assigned $B-L$ charges of  $-1$, $1/2$, and $(-4,-4,5)$, respectively, under the alternative gauged $U(1)_{B-L}$ symmetry.
\begin{widetext}
\begin{center} 
\begin{table}[t]
\begin{tabular}{|c||c|c|c|c|c||c|c|c|}\hline\hline  
Fermions& ~$Q_L$~ & ~$u_R$~ & ~$d_R$~ &~$L_L$~ & ~$e_R$~ & ~$N$~ & ~$\psi$~ & ~$\nu_{R_{1,2,3}}$~ 
\\\hline 
$SU(3)_C$ & $\bm{3}$  & $\bm{3}$  & $\bm{3}$  & $\bm{1}$  & $\bm{1}$  & $\bm{1}$  & $\bm{1}$  & $\bm{1}$  \\\hline 
 $SU(2)_L$ & $\bm{2}$  & $\bm{1}$  & $\bm{1}$ & $\bm{2}$ & $\bm{1}$  & $\bm{1}$ & $\bm{1}$ & $\bm{1}$   \\\hline 
$U(1)_Y$ & $\frac16$ & $\frac23$  & $-\frac{1}{3}$ & $-\frac12$  & $-1$ & $0$ & $0$ & $0$    \\\hline
 $U(1)_{B-L}$ & $\frac13$ & $\frac13$  & $\frac13$ & $-1$  & $-1$   & $-1$   & $\frac12$   & $(-4,-4,5)$   \\\hline
\end{tabular}
\begin{tabular}{|c||c|c|c|c|}\hline\hline
  Bosons  &~ $H$ ~ &~ $\varphi$~ &~ $\chi$ &~ $\chi'$ \\\hline
$SU(3)_C$ & $\bm{1}$   & $\bm{1}$  & $\bm{1}$ & $\bm{1}$ \\\hline 
$SU(2)_L$ & $\bm{2}$   & $\bm{1}$ & $\bm{1}$ & $\bm{1}$  \\\hline 
$U(1)_Y$ & $\frac12$  & $0$ & $0$ & $0$    \\\hline
 $U(1)_{B-L}$ & $0$ & $1$ & $\frac32$  & $\frac72$  \\\hline
\end{tabular}%
\caption{Field contents of fermions and bosons
and their charge assignments under $SU(3)_C\times SU(2)_L\times U(1)_Y\times U(1)_{B-L}$, where flavor indices are abbreviated.}
\label{tab:1}
\end{table}
\end{center}
\end{widetext}
In the bosonic sector, we add three kinds of single bosons $\varphi$, $\chi$, and $\chi'$ with $1$, $3/2$, and $7/2$ $U(1)_{B-L}$ charges, respectively. Here $\varphi$ develops a nonzero VEV $v'$, while $\chi$ and $\chi'$ are supposed to be inert bosons.
The SM Higgs is denoted by $H$ and its VEV is defined by $\langle H\rangle\equiv [0,v_H/\sqrt2]^T$.
Under these symmetries, the renormalizable Lagrangian for the lepton sector is given by 
\begin{align}
-{\cal L}_{L}&=
 y^\ell_{ii} \bar L_{L_i} e_{R_i} H + y^N_{ia}\bar L_{L_i} \tilde H N_{R_a} +M_{N_{aa}} \overline{N_{L_a} }N_{R_a}
 +M_{\psi_{\alpha\alpha}} \overline{\psi_{L_\alpha}}\psi_{R_\alpha}  \\
&
+f_{a\alpha} \overline {N_{L_a}} \psi_{R_\alpha} \chi^*+ g_{\alpha \rho} \overline{\psi^C_{R_\alpha}}  \nu_{R_\rho} \chi'
+ h_{i3} \overline{\nu^C_{R_i}} \nu_{R_3} \varphi^* 
+y^\varphi_{R_{\alpha\beta}} \overline{\psi_{R_\alpha}} \psi^C_{R_\beta}\varphi 
+y^\varphi_{L_{\alpha\beta}} \overline{\psi^C_{L_\alpha}} \psi_{L_\beta}\varphi^* 
+{\rm c.c.},\nn
\end{align}
where $\tilde H \equiv (i \sigma_2) H^*$ with $\sigma_2$ being the second Pauli matrix, the indices ($i,a,\alpha,\beta$) run over $1$ to $3$ while $\rho=1,2$, and $y^\ell,\ M_N,\ M_\psi$ are diagonal without loss of generality.
Therefore, the charged-lepton mass eigenvalues  are given by $(m_e,m_\mu,m_\tau)= \frac{v_H}{\sqrt2}(y^\ell_{11},y^\ell_{22},y^\ell_{33})$.

\subsection{ Scalar sector}
\noindent The scalar particles are parameterized as 
\begin{align}
&H =\left[\begin{array}{c}
w^+\\
\frac{v_H + h +i z}{\sqrt2}
\end{array}\right],\quad 
\varphi=
\frac{v' + h' + i z' }{\sqrt2} ,
\label{component}
\end{align}
where $w^+$ and $z$ respectively provide nonzero masses for the SM gauge bosons $W^+$and $Z$, 
and $z'$ supplies the mass for $B-L$ gauge boson $Z'$ that is simply given by $m_{Z'}\simeq g' v'$.
%
A relevant nontrivial Higgs potential is found as
\begin{align}
\frac{\lambda_0}2 \chi'\chi^* (\varphi^*)^2 +{\rm c.c.},
\end{align}
which is invariant under our charge assignments. 
This term plays a role in constructing the $\overline{N_L}\nu_R$ term at the one-loop level together with Yukawa terms.

\subsection{Neutral Fermion Sector}
\noindent After the spontaneous symmetry breaking of $H$ and $\varphi$, the mass terms for neutral fermion sector are given by
\begin{align}
&m_{D_{ia}} \overline{\nu_{L_i}} N_{R_a} 
 +(M_{\psi_R})_{\alpha\beta} \overline{\psi_{R_\alpha}} \psi^C_{R_\beta} 
+ (M_{\psi_L})_{\alpha\beta} \overline{\psi^C_{L_\alpha}} \psi_{L_\beta} 
+ M_{R_{\rho 3}} \overline{\nu_{R_\rho}} \nu_{R_3} 
 + \delta \mu_{a\rho} \overline{N_{L_a}} \nu_{R_\rho}
 \\
&  +M_{N_{aa}} \overline{N_{L_a}} N_{R_a}
 +M_{\psi_{\alpha\alpha}} \overline{\psi_{L_\alpha}}\psi_{R_\alpha}
 +{\rm h.c.} ,
\end{align}
where $m_{D_{ia}}\equiv y^N_{ia} v_H/\sqrt2$, $(M_{\psi_R})_{\alpha\beta}  \equiv  y^\varphi_{R_{\alpha\beta}}v'/\sqrt2 $,
 $(M_{\psi_L})_{\alpha\beta}  \equiv  y^\varphi_{L_{\alpha\beta}}v'/\sqrt2$,
 and 
\begin{align}
M_R=\frac{v'}{\sqrt2}
\left[\begin{array}{ccc}
0 & 0  & h_{13}   \\
0 & 0  &  h_{23}   \\
 h_{13} &  h_{23}  & 0   \\
\end{array}\right]
\equiv 
\left[\begin{array}{ccc}
0 & 0  & m_1   \\
0 & 0  &  m_2 \\
m_1& m_2  & 0   \\
\end{array}\right],
\label{eq:NR}
\end{align}
in basis of $[\nu_{R_1},\nu_{R_2},\nu_{R_3}]^T$. Note that $M_R$ is a rank-two mass matrix.
~\footnote{Consequently, the resulting light neutrino mass matrix $m_\nu$ also has a rank-two mass matrix. 
This leads to a testable prediction: the lightest neutrino mass is strictly zero
($D_1=0$ for normal hierarchy (NH), or $D_3=0$ for inverted hierarchy (IH)), 
which yields a specific prediction for the sum of neutrino masses
$\sum_i D_i$ accessible to future cosmological observations.}
Therefore, the inverse of $M_R$ cannot be defined in the  usual way.
Thus, we apply ``Moore-Penrose Pseudo Inverse Matrix", which is given by
\begin{align}
M_R^{-1}&\approx  \frac{1}{|m_2|\sqrt{1+k^2}} U_R {\rm diag}[0,-1,1] U_R^T \equiv  \frac{1}{|m_2|\sqrt{1+k^2}} \tilde M_R^{-1},\\
U_R &= \frac1{\sqrt2}
\left[\begin{array}{ccc}
-\frac{\sqrt2}{\sqrt{1+k^2}} & -\frac{k}{\sqrt{1+k^2}}  & \frac{k}{\sqrt{1+k^2}}  \\
\frac{\sqrt2 k}{\sqrt{1+k^2}}  & - \frac{1}{\sqrt{1+k^2}}   &  \frac{1}{\sqrt{1+k^2}}   \\
0&  1  & 1   \\
\end{array}\right],
\label{eq-mr_inv}
\end{align}
where $k\equiv m_1/m_2$.

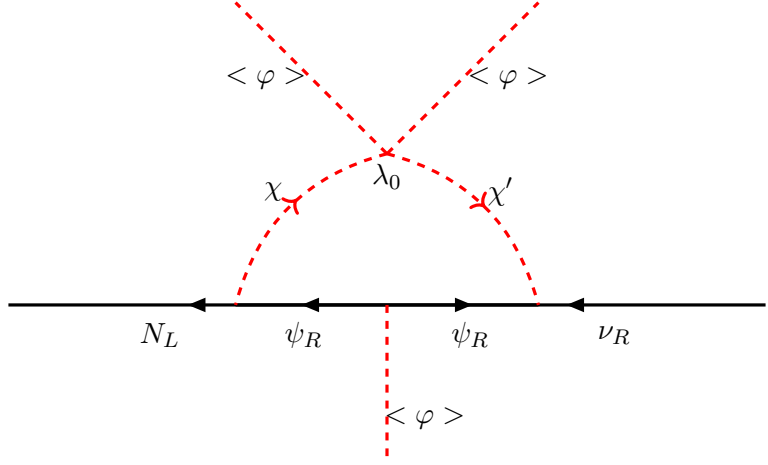
\begin{figure}[t]
    \centering
    \begin{tikzpicture}
    \begin{feynman}

    \vertex at (2,0) (i1);
    \vertex at (-2,0) (i2);
    \vertex at (0,0) (a);
    \vertex at (0, 2) (d);
    \vertex at (0,-2) (e);
    \vertex at (5,0) (b);
    \vertex at (-5,0) (c);

    \vertex at (1.1,-0.4) () {\(\psi_R\)};
    \vertex at (-1.1,-0.4) () {\(\psi_R\)};
    \vertex at (0.5,-1.5) () {\(<\varphi>\)};

    \vertex at (2,4) (f);
    \vertex at (-2,4) (g);

    \vertex at (1.6,3) () {\(<\varphi>\)};
    \vertex at (-1.6,3) () {\(<\varphi>\)};

    \vertex at (1.5,1.5) () {\(\chi'\)};
    \vertex at (-1.5,1.5) () {\(\chi\)};
    \vertex at (0,1.7) () {\(\lambda_0\)};

    \vertex at (-3,-0.4) () {\(N_{L}\)};
    \vertex at (3,-0.4) () {\(\nu_{R}\)};

    \diagram*{
        (a) -- [fermion, very thick] (i1),
        (a) -- [fermion, very thick] (i2),

        (b) -- [fermion, very thick] (a),
        (a) -- [fermion, very thick] (c),

      (d) -- [red, scalar, very thick, bend left, postaction={decorate},
decoration={markings,mark=at position 0.5 with {\arrow{>}}}] (i1),
        (d) -- [red, scalar, very thick, bend right, postaction={decorate},
decoration={markings,mark=at position 0.5 with {\arrow{<}}}] (i2),

        (a) -- [red, scalar, very thick] (e),

        (d) -- [red, scalar, very thick] (f),
        (d) -- [red, scalar, very thick] (g),
    };

    \end{feynman}
    \end{tikzpicture}

    \caption{Feynman diagram for radiative generation of $\delta\mu$.}
    \label{fig:diag}
\end{figure}
\noindent The term $\delta \mu_{a\rho}$ is obtained radiatively at the one-loop level as shown by the Feynman diagram given in Fig.~\ref{fig:diag} and its form is found as
\begin{align}
\delta \mu_{a\rho}&
\simeq- \frac{\lambda_0 v'^2} {2(4\pi)^2} \frac{f_{a\alpha} M_{\psi_{R_\alpha}} g_{\alpha \rho}}{m^2_\chi-M_{\psi_{R_\alpha}}^2}
\left[
\frac{m_\chi^2}{m_\chi^2- M_{\psi_{R_\alpha}}^2}\ln\left(\frac{m_\chi^2}{m_{\chi'}^2}\right)
-\frac{m_{\chi'}^2}{m_{\chi'}^2- M_{\psi_{R_\alpha}}^2}\ln\left(\frac{m_{\chi'}^2}{m_\chi^2}\right)
\right]\\
&\equiv - \frac{\lambda_0 v'} {\sqrt2(4\pi)^2} {\delta\tilde\mu_{a\rho}}
,
\end{align}
where we assume $M_{\psi_{R_{\alpha\beta}}}\approx M_{\psi_{R_{\alpha\alpha}}}\delta_{\alpha\beta}$ for simplicity,
and applied the mass insertion approximation assuming $\lambda_0 v'^2\ll m_\chi^2, \ m_{\chi'}^2$.
Note that $m_\chi^2$ and $ m_{\chi'}^2$ respectively represent the mass-squared eigenvalues for $\chi$ and $\chi'$.

\noindent The neutral fermionic mass matrix in basis of $\nu^C_L,N_R, N^C_L,\nu_R$ is given by
\begin{align}
\left[\begin{array}{cccc}
0 & m_D& 0& 0   \\
m_D^T & 0 & M_N^T  & 0 \\
0 & M_N  & 0& \delta\mu   \\
0 & 0  & \delta\mu & M_R  \\
\end{array}\right].
\label{eq-Nmass}
\end{align}
Then, considering the following mass hierarchies 
\begin{align}
\delta\mu\ll  m_D < M_N, M_R,
\label{eq:hierarchy}
\end{align}
the active neutrino mass matrix is given by
\begin{align}
m_\nu &\approx -(m_D M_N^{-1}) (\delta\mu M_R^{-1}\delta\mu^T) (m_D M_N^{-1})^T\\
& = -\frac{\lambda_0^2}{1024\pi^4} \frac{v_H^2 v'^2}{M^2_3 |m_2|\sqrt{1+k^2}} 
(y^N  \tilde M_N^{-1}) (\delta\tilde\mu \tilde M_R^{-1}\delta\tilde\mu^T) (y^N \tilde M_N^{-1}) ^T\\
&
=\kappa_\nu \tilde m_\nu
,
\label{eq-Nmass}
\end{align}
where $M_N={\rm diag}[M_1,M_2,M_3]\equiv M_3{\rm diag}[\tilde M_1,\tilde M_2,1] \equiv M_3\tilde M_N$. This form is called ``double inverse seesaw".
The $m_\nu$ can be diagonalized by $D_\nu(\equiv \kappa_\nu \tilde D_\nu) = U^\dag_\nu m_\nu U^*_\nu(\equiv \kappa_\nu U^\dag_\nu\tilde m_\nu U^*_\nu)$, where we can identify $U_\nu = U^{\rm PMNS} P$ with $P={\rm diag}[1,e^{i\alpha_{21}/2},1]$ without loss of generality since the charged-lepton mass matrix is diagonal. In the standard parametrization of $U^{\rm PMNS}$ in Particle Data Group~\cite{ParticleDataGroup:2026mpi}, three mixing angles, $s^2_{13},\ s^2_{23},\ s^2_{12}$ and two phases, Dirac CP phase $\delta_{\rm CP}$ and Majorana phase  $\alpha_{21}$ are respectively given by
\begin{align}
& s^2_{13} =|U^{\rm PMNS}_{e3}|^2,\quad s^2_{23} =\frac{|U^{\rm PMNS}_{\mu3}|^2}{1-|U^{\rm PMNS}_{e3}|^2},\quad 
s^2_{12} =\frac{|U^{\rm PMNS}_{e2}|^2}{1-|U^{\rm PMNS}_{e3}|^2},\\
& \delta_{\rm CP} = -{\rm arg}[U^{\rm PMNS}_{e3}],\quad \alpha_{21} = 2 {\rm arg}[U^{\rm PMNS}_{e2}]. 
\end{align}
$\kappa_\nu$ is given by an observable $\Delta m^2_{\rm atm}$ and dimensionless mass eigenvalues.
Then, the other observables $\Delta m^2_{\rm sol}$ and sum of the neutrino masses can be fitted in term of $\Delta m^2_{\rm atm}$ and dimensionless mass eigenvalues,
depending on the neutrino mass ordering;
\begin{align}
& ({\rm NH}):\ D_1=0,\  \kappa_\nu^2 = \frac{\Delta m^2_{\rm atm}}{\tilde D_3^2},\quad \Delta m^2_{\rm sol}= \frac{\tilde D_2^2}{\tilde D_3^2}\Delta m^2_{\rm atm},
\quad \sum_i D_i \simeq \sqrt{\Delta m^2_{\rm atm} } \sim 60 \ {\rm meV},
\\
& ({\rm IH}):\ D_3=0,\  \kappa_\nu^2 = \frac{\Delta m^2_{\rm atm}}{\tilde D_2^2},\quad \Delta m^2_{\rm sol}= \left(1-\frac{\tilde D_1^2}{\tilde D_2^2}\right)
\Delta m^2_{\rm atm},\quad \sum_i D_i \simeq 2\sqrt{\Delta m^2_{\rm atm}} \sim 120 \ {\rm meV}.
\end{align}
Here, we discuss how natural the mass hierarchies in Eq.~(\ref{eq:hierarchy}) are.
Since $\delta \mu$ is one-loop induced, this is naturally smaller than the other masses.
The order of $m_D$ is essentially given by $v_H$ after the electroweak spontaneous symmetry breaking, its scale would be expected to be 100 GeV.
On the other hand, the order of $M_{\psi_R}$ is provided by the new physics (NP) scale after the $B-L$ spontaneous symmetry breaking, its scale would be more than 1 TeV.
The scale of $M_N$ is in principle arbitrary, since the mass is directly induced due to the Dirac feature. However, we simply suppose that 
$M_N$ also originates from the NP scale, therefore we expect the order of $M_N$ to be greater than 1 TeV.
Thus, the hierarchies among them are reasonably explained to some extent.

\noindent The non-unitarity constraints restrict the mass ratio between $m_D$ and $M_N$ in the following ways depending on normal hierarchy (NH) and inverted hierarchy (IH)~\cite{Blennow:2023mqx}:
\begin{align}
&({\rm NH}):\quad
|m_D^* (M_N^*)^{-1} (M_N^T)^{-1} m_D^T |\lesssim
\left[\begin{array}{ccc}
1.78\times10^{-5} & 1.66\times10^{-5}  & 3.0\times10^{-5}   \\
1.66\times10^{-5} & 1.38\times10^{-4}  & 1.44\times10^{-4}   \\
3.0\times10^{-5} & 1.44\times10^{-4}  & 1.72\times10^{-4}   \\
\end{array}\right],\\
&({\rm IH}):\quad
|m_D^* (M_N^*)^{-1} (M_N^T)^{-1} m_D^T |\lesssim
\left[\begin{array}{ccc}
1.96\times10^{-4} & 7.4\times10^{-6}  & 5.0\times10^{-5}   \\
7.4\times10^{-6} & 4.0\times10^{-7}  & 7.6\times10^{-7}   \\
5.0\times10^{-5} &  7.6\times10^{-7}  & 1.88\times10^{-5}   \\
\end{array}\right]. 
\label{eq:nonunitarity}
\end{align}
These constraints lead us to the following bounds:
\begin{align}
&({\rm NH}):\quad
4.07\times 10^{-3}\lesssim
|m_D/M_N|
\lesssim
1.31\times 10^{-2},
\\
&({\rm IH}):\quad
6.32\times 10^{-4}\lesssim
|m_D/M_N|
\lesssim
1.40\times 10^{-2}. 
\label{eq:bounds}
\end{align}
These experimental bounds are consistent with our theoretical expectation. 

Although the specific structure of $M_R$ leads to the particular feature of a rank-two neutrino mass matrix, the matrices $\delta\mu$ and $y^N$ possess full degrees of freedom as $3\times3$ matrices, making it clear that a valid solution exists. Furthermore, since the primary focus of our paper is to elucidate the origin of the mass scales (orders of magnitude) for each mass matrix, we do not explicitly perform a numerical analysis to fit the neutrino oscillation data.

\section{Dark matter candidates}
\label{sec:dm}
\noindent Here, we discuss our DM candidate. In our model, there exist two DM candidates: the lightest neutral fermion of $\psi$ or the lightest inert boson of $\chi^{(')}$. In our analysis, we rely on a Breit-Wigner enhancement in the s-channel process via $Z'$ gauge boson to explain the relic density of DM.
 Therefore, we suppose the other processes such as Higgs portals are sub-dominant.
 
\subsection{Fermionic DM}
\noindent First, let us discuss the fermionic DM candidate $X\equiv \psi_1$, where we assume the DM to be a Dirac fermion assuming $M_{\psi_L,\psi_R}<<M_{\psi}$.
Therefore, the DM mass is $m_X\equiv M_{\psi_{11}}$. 
Then, the relevant Lagrangian to describe the relic density is given by
\begin{align}
g' Q_X \bar X \gamma^\mu X Z'_\mu 
+
g' \sum_{f=SM}Q_f \bar f \gamma^\mu f Z'_\mu ,
\end{align}
where $g'$ is the B-L gauge coupling, $Q_X(=1/2)$ is the B-L charge of $X$, $Q_f$ is the B-L charge of the SM fermions. 
Note that we assume the final-pair states of $\nu_R$ and $\varphi$ are kinematically forbidden, {\it i.e.}, $m_X < M_N, M_R, m_\varphi$.
The thermally averaged cross section is given by
\begin{align}
& \langle \sigma v \rangle \simeq 3Q_X^2\sqrt{\pi} x^{3/2} \left(\frac{S_{\rm anni}}{S_{\rm decay}}\right) \frac{g'^2}{m^2_{Z'}} \sqrt{\delta} e^{-\frac{x}{4}\delta},\\
& \delta\equiv \frac{m^2_{Z'} -4 m_X^2}{m_X^2},
\quad
S_{\rm anni} \approx \sum_{f_{\rm anni}} N^f_C Q^2_{f_{\rm anni}},
\quad
S_{\rm decay} \approx  \sum_{f_{\rm decay}}  N^{f_{\rm decay}}_C Q^2_{f_{\rm decay}},
\end{align}
where $f_{\rm anni}$ and $f_{\rm decay}$ are respectively the final states of particle-pairs in processes of the DM annihilation and the $Z'$ decay.  
In our case, $f$ runs over only the SM fermions and $S_{\rm anni} \sim S_{\rm decay}$.~\footnote{Even though we fix $m^2_{Z'}>2 m_X$ to pick up the resonant pole, the decay width of $Z'\to X\bar X$ can be neglected compared to the decay width of $Z'\to f\bar f$ since the decay width of  $Z'\to X\bar X$ goes to zero at the resonant point at $m^2_{Z'}=2 m_X $.}
Furthermore, we have used a narrow width approximation to obtain the above relation, therefore we have assumed to be $\Gamma_{Z'}/m_{Z'}\ll1$.
\begin{align} 
\langle \sigma v \rangle_{\rm max} &\approx 4.56 Q_X^2 x  \left(\frac{S_{\rm anni}}{S_{\rm decay}}\right) \frac{g'^2}{m^2_{Z'}}
 \sim 0.28 x  \frac{g'^2}{m^2_{X}},
\end{align}
where we have used $m^2_{Z'}\approx 4 m^2_X(1+\frac14\delta_{\rm opt})\approx 4 m^2_X(1+\frac2{x})$.
Assuming that  our maximal cross section can saturate the observed cross section $\langle \sigma v \rangle_{\rm max}=1.9\times10^{-9}$ GeV$^{-2}$ to satisfy the relic density, 
the required B-L gauge coupling $g'_{\rm relic} $ is given by
\begin{align} 
g'_{\rm relic} &
 \sim 
 1.65\times10^{-5}  \left(\frac{25}{x}\right)^{1/2}  \left(\frac{m_X}{\rm GeV}\right).
\end{align}
Collider physics at LHC provides the following constraint 
\begin{align}
\frac{g'}{m_{Z'}}\lesssim \frac{1}{7000 \ {\rm GeV}} =  \frac{1.43 \times10^{-4}}{\rm GeV} .
\end{align}
Our case at the resonant point $m_{Z'}=2 m_X$ leads to the following relation
\begin{align}
\frac{g'_{\rm relic} }{m_{Z'}} \sim \frac{g'_{\rm relic} }{2 m_{X}} \sim \frac{8.6\times10^{-6}}{\rm GeV} <  \frac{1.43 \times10^{-4}}{\rm GeV} .
\end{align}
Thus, the collider constraint is satisfied.
\\

\noindent Next, we need to check the constraint of direct detection via $Z'$ boson portal.
The scattering cross section at the resonant point is given by
\begin{align}
\sigma^N_{\rm SI} \approx \frac{\mu^2_{XN}}{\pi}  \left(\frac{g'^2}{2 m^2_{Z'}}\right)^2  \left(\frac{25}{x}\right)^{2},
\end{align}
where $\mu_{XN}\equiv m_N m_X/(m_N+m_X)$.
Inserting $g'=g'_{\rm relic}$ into the above equation, we find that
\begin{align}
\sigma^N_{\rm SI} \approx 1.27\times 10^{-49} {\rm cm^2}   \left(\frac{\mu_{XN}}{0.94{\rm GeV}}\right)^2  \left(\frac{25}{x}\right)^{2} .
\end{align}
Assuming $m_N \ll m_X$, we have $\mu_{NX}\sim m_N\sim$0.94 GeV.
The most stringent upper bound is provided by the LUX-ZEPLIN (LZ) experiment~\cite{LZ:2024zvo}, which is
$2.2\times10^{-48}$ cm$^2$ for a mass of 40 GeV at 90 \% confidence level. 
From the above relation, the direct detection upper bound is also satisfied in case of $x=25$.

 \subsection{Bosonic DM}
\noindent Second, let us move on to discuss the bosonic DM candidate $\chi_1$ or $\chi_2$, where we suppose the DM to be $X\equiv \chi_1$.
Therefore, the DM mass is $m_X\equiv m_{\chi_{1}}$. 
Then, the relevant Lagrangian to describe the relic density is given by
\begin{align}
i g' Q_X (X^*\partial^\mu X - X\partial^\mu X^*) Z'_\mu 
+
g' \sum_{f=SM}Q_f \bar f \gamma^\mu f Z'_\mu ,
\end{align}
where $Q_X(=3/2)$ is the B-L charge of $X$.
We assume final states are the same as the case of FDM. 
Similar to the case of FDM, the thermally averaged cross section is given by
\begin{align}
& \langle \sigma v \rangle \simeq g'^2 Q_X^2 \frac{S_{\rm anni} }{24m_X^2\sqrt{\pi}} \frac{x^{3/2}}\gamma \delta^{3/2} e^{-\frac x4\delta},
\end{align}
where $\gamma\sim\frac{g'^2 m^2_{Z'}}{12\pi m_X^2}S_{\rm decay}$.
It is straightforwardly to find that $\langle \sigma v \rangle$ becomes maximum at $\delta_{\rm opt} = 6/x$, by maximizing $\delta^{3/2} e^{-\frac{x}{4}\delta}$. Therefore 
\begin{align} 
\langle \sigma v \rangle_{\rm max} & \approx 0.683 Q^2_X \left(\frac{S_{\rm anni}}{S_{\rm decay}}\right) \frac{g'^2}{m^2_X}
 \sim 1.54\frac{g'^2}{m^2_{X}},
\end{align}
where $S_{\rm anni}\sim S_{\rm decay}$ and their values are $5$ when the SM contributions are taken into account.
Assuming that  our maximal cross section can saturate the observed cross section $\langle \sigma v \rangle_{\rm max}=1.9\times10^{-9}$ GeV$^{-2}$ to satisfy the relic density, 
the required B-L gauge coupling $g'_{\rm relic} $ is given by
\begin{align} 
g'_{\rm relic} &
 \sim 
 3.52 \times10^{-5}  \left(\frac{m_X}{\rm GeV}\right).
\end{align}
Note here that the $g'_{\rm relic} $ in case of BDM does not depend on $x$, unlike the FDM case.
Furthermore, the required $g'_{\rm relic} $ in case of BDM is twice as large as that in the FDM case for $x=25$. 

\noindent Collider physics at LHC provides the following constraint 
\begin{align}
\frac{g'}{m_{Z'}}\lesssim \frac{1}{7000 \ {\rm GeV}} =  \frac{1.43 \times10^{-4}}{\rm GeV} .
\end{align}
Our case at the resonant point $m_{Z'}=2 m_X$ leads to the following relation
\begin{align}
\frac{g'_{\rm relic} }{m_{Z'}} \sim \frac{g'_{\rm relic} }{2 m_{X}} \sim \frac{1.85\times10^{-5}}{\rm GeV} <  \frac{1.43 \times10^{-4}}{\rm GeV} .
\end{align}
Thus, the collider constraint is satisfied.
\\

\noindent For BDM, the direct detection constraint via the $Z'$ portal is determined by
the scattering cross section at the resonant point, and its form is given by
\begin{align}
\sigma^N_{\rm SI} \approx \frac{\mu^2_{XN}}{\pi}  \left(\frac{3g'^2}{2 m^2_{Z'}}\right)^2
\end{align}
where $\mu_{XN}\equiv m_N m_X/(m_N+m_X)$.
Inserting $g'=g'_{\rm relic}$ into the above equation, we find that
\begin{align}
\sigma^N_{\rm SI} \approx 2.07\times 10^{-47}{\rm cm^2}   \left(\frac{\mu_{XN}}{0.94{\rm GeV}}\right)^2 .
\end{align}
This scattering cross section is ruled out by  the LUX-ZEPLIN (LZ) experiment.
In addition, PandaX-4T experiment~\cite{PandaX:2024qfu} provides the upper bound $1.6\times10^{-47}$ cm$^2$ for a mass of 40 GeV at 90 \% confidence level,
XENONnT experiment~\cite{XENON:2025vwd} provides the upper bound $1.7\times10^{-47}$ cm$^2$ for a mass of 30 GeV at 90 \% confidence level.
These experiments also disfavor the BDM scenario via the $Z'$ portal. Thus, 
additional interactions, such as Higgs portal couplings, would be required to make the BDM scenario viable.

\section{Conclusion}
\label{sec:con}
\noindent In this paper, we have proposed a radiative double inverse seesaw model based on an alternative gauged $U(1)_{B-L}$ symmetry. The non-universal $B-L$ charge assignment of $(-4, -4, 5)$ for the right-handed neutrinos naturally suppresses the tree-level mass generation, leading to the radiative mass generation at the one-loop level. This setup not only provides a concrete realization of the 't Hooft naturalness but also enriches the particle content with vector-like fermions and inert singlet scalars. The spontaneous breaking of the $U(1)_{B-L}$ symmetry induces a remnant $Z_2$ symmetry, rendering the lightest neutral fermion and the inert scalars as viable dark matter (DM) candidates.

\noindent We have evaluated the phenomenological viability of both fermionic and bosonic DM candidates, focusing on the $Z'$ boson portal. We found that the fermionic DM can successfully account for the observed relic density while satisfying the stringent constraints from direct detection experiments (e.g., LZ) and collider bounds. On the other hand, the bosonic DM scenario via the $Z'$ portal yields a spin-independent scattering cross section that exceeds the current upper bounds from the LZ, PandaX-4T, and XENONnT experiments, indicating that additional interactions, such as the Higgs portal, would be necessary to make it viable. 

\noindent Consequently, the fermionic DM is highly favored in our model setup. Our framework offers a testable and natural explanation for the neutrino mass hierarchy and the dark matter relic abundance. Future work could extend this analysis to include the Higgs portal effects for the bosonic DM, as well as detailed collider signatures of the vector-like fermions and the $Z'$ boson at the LHC and future experiments.

\begin{acknowledgments}
\noindent HO is supported by Zhongyuan Talent (Talent Recruitment Series) Foreign Experts Project. 
\end{acknowledgments}

\bibliography{ctma4.bib}

\end{document}